\documentclass[useAMS,usenatbib,usegraphicx]{mn2e}  

\usepackage{graphicx,breqn}
\DeclareGraphicsExtensions{.pdf,.png,.jpg,.ps}
\usepackage{epsfig}
\usepackage{color}

\begin{document}

%%%%%%%%%%%%%%%%%%% MNRAS %%%%%%%%%%%%%%%%%%%%

\title[The Void in the Sculptor Group Spiral Galaxy NGC 247]{The Void in the Sculptor Group Spiral Galaxy NGC 247}

\author[R. Wagner-Kaiser, T. DeMaio, A. Sarajedini, S. Chakrabarti]
  {R.~Wagner-Kaiser$^1$, 
  T.~Demaio$^1$, 
  A.~Sarajedini$^1$, 
  S.~Chakrabarti$^2$
\\
  $^1$University of Florida, Department of Astronomy, 211 Bryant Space Science Center, Gainesville, FL, 32611\\
  $^2$Rochester Institute of Technology, School of Physics and Astronomy, 84 Lomb Memorial Drive, Rochester, NY 14623\\
}

\date{}

\pagerange{\pageref{firstpage}--\pageref{lastpage}} \pubyear{2014}

\maketitle

\label{firstpage}

\begin{abstract}
The dwarf galaxy NGC 247, located in the Sculptor Filament, displays an apparent void on the north side of its spiral disk. The existence of the void in the disk of this dwarf galaxy has been known for some time, but the exact nature and cause of this strange feature has remained unclear. We investigate the properties of the void in the disk of NGC 247 using photometry of archival Hubble Space Telescope data to analyze the stars in and around this region. Based on a grid of isochrones from log(t)=6.8 to log(t)=10.0, we assign ages using nearest-neighbor interpolation. Examination of the spatial variation of these ages across the galaxy reveals an age difference between stars located inside the void region and stars located outside this region. We speculate that the void in NGC 247's stellar disk may be due to a recent interaction with a nearly dark sub-halo that collided with the disk and could account for the long-lived nature of the void.
\end{abstract}

\maketitle


\vspace{2pc}

%\maketitle

%%%%%%%%%%%%%%%%%%%%%%%%%%%%%%%%%%%%%%%%%%%  Introduction

\section{Introduction}\label{Intro}

The Sculptor group has been shown to be a dynamically unevolved filament of galaxies, the ends of which are not gravitationally bound but are instead moving with the Hubble flow (Karachentsev et al. 2003). With a different density and population of galaxies than our own Milky Way and companions, studying the Sculptor filament galaxies holds the promise of new insights into galaxy formation and evolution. NGC 247 and its closest companion, NGC 253, make up the core of the Sculptor filament.

NGC 247 is a dwarf spiral galaxy (SAB(s)d) that is perhaps best known simply for being the companion to NGC 253. Closer inspection of NGC 247 shows an apparent void in its disk between the nucleus and outer arm on the north side of the galaxy. This void is visible in bandpasses from the infrared to the ultraviolet and thus far has drawn little attention in the literature beyond noting its existence (Roberts 1915, Carignan 1985). At a distance of about 3.5 Mpc, the void region in NGC 247 is approximately 3.4 kpc long. For a dwarf galaxy, this makes up a significant percentage of its stellar disk, which is 11.3 kpc in extent along the major axis (Jarrett et al. 2003).

Several ground-based studies have observed the central, outer, and extra-planar regions of the galaxy (Davidge 2000; Davidge 2006). These studies find an extended stellar disk with young stars forming beyond the HI region, another unusual aspect of NGC 247. Observations of NGC 247 were also obtained with the Hubble Space Telescope (HST) as part of the ANGST survey (Dalcanton et al. 2009). However, neither of these particular sets of observations included the void region of the galaxy.

We note that the void is a significantly larger ``bubble" than would be expected from any typical, singular supernovae. A similar bubble was observed in NGC 1620, an SBbc spiral galaxy (Chaboyer \& Vader 1991; Vader \& Chaboyer 1995). It was determined that this bubble could have been caused by many concurrent supernovae, pushing material outwards in a ring or arc up to a couple of kiloparsecs in size. Vader \& Chaboyer (1995) concluded that the more than 150 supernovae necessary to create the bubble in NGC 1620 originated from a super-luminous cluster visible in the center of the ring. In addition, Efremov et al. (1998) suggest that powerful gamma ray bursts (GRBs) may be responsible for creating large holes in galactic disks; this scenario would lessen the need for large numbers of supernovae. They also posit that any super-luminous clusters may fade in a short timescale so that they are no longer visible inside of the bubble, meaning that while a bubble may remain, the visual evidence of the stars causing it may no longer be present.

Larsen (1999) identified three young, massive star clusters in NGC 247. One of these clusters, referred to as n247-899 by Larsen (1999), is located near the southern edge of the void. Being a more massive cluster, it is plausible that this cluster could have harbored the necessary amount of energy to produce the void that is seen, but it is unclear without further examination.

Previous studies have investigated the X-ray halo of NGC 247, the results of which suggest that there may be material falling into the inner regions of the galaxy (Strassle et al. 1999). However, there has been no trace of recent star formation within approximately the past 70 Myr, despite the possibility of infalling material (Gordon et al. 1999; Davidge 2000). Recent studies of X-ray sources include observations from XMM-Newton, Chandra, and optical imaging with HST (Jin et al. 2011; Tao et al. 2012). These studies show a supersoft ultra-luminous X-ray source and optical counterpart with a large stellar association nearby on the western side of NGC 247, near the nuclear region. The lack of an X-ray signature associated with or close to the void region suggests that neither supernovae nor GRBs are likely to have caused the void in NGC 247.

Early radio studies found that there is twice as much HI in NGC 247 in the northern half of the galaxy as in the southern portion (Carignan 1985; Carignan \& Puche 1990). These studies also find that the HI envelope of the galaxy is relatively compact, truncating before the stellar disk, a matter further discussed by Davidge (2006). More recent studies have mapped out the HI at higher resolution (Warren et al. 2012; Ott et al. 2012), showing NGC 247 to have patchy HI throughout its disk. If the void is a region lacking in star formation, we would expect to find less HI emission in that region. Previous studies have not compared the void region and the HI disk characteristics; we will examine this herein (see Section \ref{Star Formation Tracers}). We will also examine the H-$\alpha$ distribution from Hlavacek-Larrondo et al. (2011) as a star formation tracer.

An area of significant extinction in the disk could potentially account for the observed appearance of the NGC 247 void. However, as noted above, the void structure is visible in many wavelengths: from ultraviolet observations by GALEX to infrared observations from the Spitzer Space Telescope (MIPS and IRAC). Thus, it seems unlikely for the void to be purely the result of strong extinction. Additionally, the HI observations (Warren et al. 2012; Ott et al. 2012) show a dip in strength in the void region, suggesting that gas and dust do not strongly populate the region, corroborating the evidence against extinction as the primary cause of the void.

Galaxy interactions may be a viable means of producing the void. Previous studies have tried to find concrete evidence for interactions between NGC 247 and NGC 253 (Whiting 1999; Davidge 2006; Davidge 2010). At a projected distance of 350 kpc from NGC 247, NGC 253 shows a compact HI distribution similar to the observed truncation in NGC 247's HI disk (Davidge et al. 2010; Carignan \& Puche 1990). This is broadly consistent with a scenario in which the disks of NGC 253 and NGC 247 have been tidally disrupted because of interactions between the two galaxies. The spin of NGC 253's disk is also consistent with torque caused by NGC 247, offering further evidence that these two systems have influenced each other's evolution (Whiting 1999). However, NGC 247 does not show any of the classical signs of interaction such as a central starburst or tidal features (Davidge et al. 2010). 

The extension of NGC 247's stellar disk beyond the HI disk may be indicative of interactions, where stars could have been cannibalized onto NGC 247's stellar disk from the interacting galaxy (Ibata et al. 2003; Erwin et al. 2005). However, to match the young stars found in the extended stellar disk of NGC 247, this interaction would have had to happen in the past 10 to 20 million years (Davidge 2006). There is a lack of a nearby companion identified that could be responsible for extending the stellar disk in the necessary timeframe. While the presence of the void could plausibly be due to interactions with NGC 253, it would be unusual to find an outer ring intact and all inner material cleared out, as we see in NGC 247. 
%No concrete evidence of interactions between the two galaxies have been presented in previous studies.

Another scenario to consider is the existence of spiral density waves. Results from simulations by Minchev et al. 2012 show that discontinuities in spiral galaxies may arise from different interacting spiral modes or the break between inner structure and outer structure in the galaxy. Such an occurrence in NGC 247 could provide the necessary conditions for a long-lived void-like structure to exist in a ``lopsided" spiral. However, these simulations are based on models with a strong bar in the galaxy, whereas NGC 247's bar is weak. Additionally, other studies suggest that dwarf galaxies overall tend to lack spiral density waves (Ott et al. 2012, Brosch et al. 1998).

One possibility is that the void was caused by the interaction of a dark sub-halo that collided with the disk in the recent past. The current paradigm of structure formation in the universe predicts a wealth of sub-structure on all scales -- from galaxy clusters, to spiral galaxies, to dwarf spirals.   From dissipationless cosmological simulations, we expect roughly one interaction with a $\sim$ 1:100 mass ratio satellite over a dynamical time (Diemand et al. 2008).  The abundance of sub-structure in cosmological simulations has prompted a number of authors to explore the effect of sub-halo interactions on galactic disks.  Chakrabarti \& Blitz (2009; 2011) explained the observed perturbations in the outer HI disk of the Milky Way using a model of a (nearly) dark sub-halo that has a close collision with the Galactic disk, and predicted the mass and location (in radius and azimuth) of this putative satellite.  Quillen et al. (2009) showed that radial mixing of stellar metallicities can be effectively accomplished in a scenario where the Milky Way has experienced a close encounter with a satellite of mass ratio $\sim$ 1:300.  Discoveries of Milky Way dark-matter dominated satellite galaxies (Belokurov et al. 2006; Wilman et al. 2006, among others), some fainter than some star clusters, suggest that there may be many more waiting to be discovered.  Some of these dim satellites may tidally interact with the Galactic disk.  

Theoretical models suggest that $\sim$ 1000 galaxies with $M_{\star} \ga 10^{3} M_{\odot}$ await discovery within approximately 3 Mpc relative to the galactic center (Garrison-Kimmel et al. 2013).  The Sculptor Group is $\sim$ 3 Mpc from the center of the Milky Way; systems like the Sculptor Group that are in relative close proximity to our own galaxy are the next frontier for the discovery of ultra-faint dwarfs.  In section \ref{Discussion}, we consider the possibility of the formation of the void in NGC 247 due to a collision with a sub-halo.  In this discussion, we emphasize the distinction between gas-bearing sub-halos and purely dark-matter dominated sub-halos, as they lead to different signatures in the galactic disk (Kannan et al. 2012).

This paper is organized as follows. In Section \ref{Data}, we discuss the HST data we use as well as our processing techniques. With deep CMDs from HST images, we can trace the physical properties of NGC 247. In Sections \ref{Colors} and \ref{Ages}, we present our analysis of the void region with respect to its stellar populations. Using isochrones, we attribute approximate ages for stars inside and outside of the void region. From this, we examine the spatial distribution of stellar ages in NGC 247 and compare it to the HI and H-$\alpha$ distributions. Section \ref{Star Formation Tracers} examines star formation tracers in NGC 247. In doing so, we gain insight into the nature and cause of this strange void in NGC 247. In Section \ref{Discussion}, we present our analysis of the formation and evolution of the void region and our conclusions are presented in Section \ref{Conclusion}.

%%%%%%%%%%%%%%%%%%%%%%%%%%%%%%%%%%%%%%%%%%%  Data

\section{Data}\label{Data}

\begin{table*}
\centering
%    \begin{minipage}{180mm}
    \caption{HST Archival Images}
    \begin{tabular}{@{}cccccc@{}}
    \hline
 \textbf{GO PI} &  \textbf{RA} & \textbf{Dec} & \textbf{Filter}& \textbf{Exposure Time} \\  
\hline
Karachentsev &	11.775417	& 	-20.651111  &	F606W  &    2 x 600  \\
    		 &	            & 	            &	F814W  &    2 x 450  \\  
        \hline
    \end{tabular}
%    \end{minipage}
   \label{Images}
\end{table*}

\subsection{Observations}

\begin{figure}
  \centering
    \includegraphics[scale=0.45]{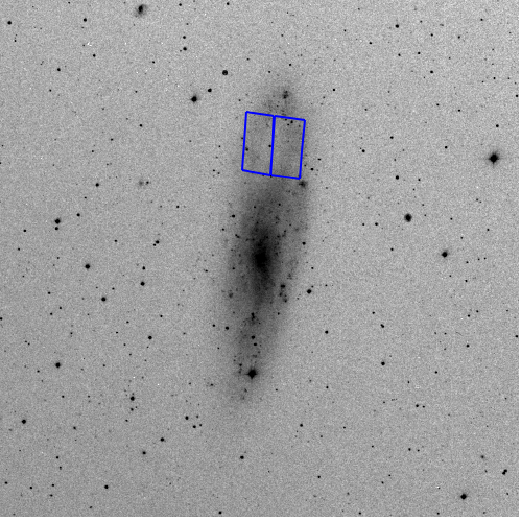}
  \caption{DSS image of NGC 247, with the HST image footprint shown, positioned over the void region.}
  \label{footprint}
\end{figure}

The archival HST images of NGC 247 are from a survey of neighboring galaxies (Cycle 12 GO proposal 9771, PI: Karachentsev) taken with the Advanced Camera for Surveys Wide Field Channel (ACS/WFC). There are two images of NGC 247 in the F606W filter, each with 600 second exposures, and two images in F814W, each with 450 second exposures. This set of frames is conveniently centered directly on the void region of NGC 247, the location of interest, as seen in Figure \ref{footprint}. Table \ref{Images} shows a log of the observations used in this study.

\subsection{Reduction}

We retrieved the FLC images of the NGC 247 field from the Mikulski Archive for Space Telescopes (MAST). These are the FLT frames corrected for the effects of charge transfer efficiency. The data quality files were applied and the resultant images were photometered as described by Sarajedini et al. (2006), using the DAOPHOT, ALLSTAR, and ALLFRAME software packages (Stetson 1987; 1994) to yield instrumental magnitudes. These were combined for the frames in each filter and then matched to form colors. We then applied the Sirianni et al. (2005) transformations to bring the magnitudes onto the VegaMAG system. 
In order to derive the highest quality photometric dataset, error cuts were applied based on the DAOPHOT Chi, Sigma, and Sharpness parameters. This process yielded 328,901 stars across the two WFC chips to be used in our subsequent analysis.

\begin{table*}
\centering
%    \begin{minipage}{180mm}
    \caption{Properties of NGC 247}
    \begin{tabular}{@{}cccccc@{}}
    \hline
\textbf{Property} &  \textbf{Value} & \textbf{Reference} \\  
\hline
Right Ascension     &   0$^{h}$47$^{m}$8.3$^{s}$  &   Dalcanton et al. 2009 \\
Declination         &   20$^{\circ}$45'36"   &   Dalcanton et al. 2009 \\
Morphological type  &	SAB(s)d     &   Konstantopoulos et al. 2013 \\
Inclination         &   75.4$^{\circ}$  &   Carignan 1985  \\
Absolute Magnitude  &   -18.28          &   Karachentsev et al. 2003  \\
Mass                &   $\sim$10$^{10}$ M$_{\odot}$ &   Strassle et al. 1999 \\
Distance            &   $\sim$3.5 - 4 Mpc   &   Various studies (Warren et al. 2012, Dalcanton et al. 2009, \\
&  &   Davidge 2006, Karachentsev et al. 2003, Strassle et al. 1999) \\
        \hline
    \end{tabular}
%    \end{minipage}
   \label{basic}
\end{table*}

%%%%%%%%%%%%%%%%%%%%%%%%%%%%%%%%%%%%%%%%%%%  Results

\section{Analysis and Results}\label{Results}

\begin{table*}
\centering
%    \begin{minipage}{180mm}
    \caption{Published Distances and Extinctions}
    \begin{tabular}{@{}cccccc@{}}
    \hline
\textbf{Distance} &  \textbf{E(B-V)} & \textbf{Method} & \textbf{Reference} \\  
\hline
28.06   & 0.19       & Tully-Fisher          &   Karachentsev et al. 2003b \\
27.81   & 0.26       & TRGB                  &   Karachentsev et al. 2006 \\
27.9    & 0.19       & TRGB (i')             &   Davidge 2007 \\
27.80   & 0.13       & PL Relation (visual)  &   Garc\'{i}a-Varela et al. 2008  \\
27.723  & 0.17       & TRGB (F475W,F814W)    &   Dalcanton et al. 2009 \\
27.745  & 0.17       & TRGB (F606W,F814W)    &   Dalcanton et al. 2009 \\
27.64   & 0.18       & PL Relation (IR)      &   Gieren et al. 2009 \\
        \hline
    \end{tabular}
%    \end{minipage}
   \label{dists}
\end{table*}

\begin{figure}
  \centering
    \includegraphics[scale=0.45]{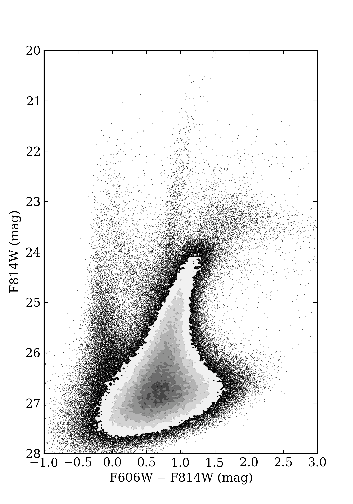}
  \caption{A CMD of the photometry we performed on the HST images of the void region from the Cycle 12 imaging of NGC 247's disk.}
  \label{CMD}
\end{figure}

Our color-magnitude diagram of the NGC 247 void field is shown in Figure \ref{CMD}. Based on the range of values published in the literature for NGC 247 (see Table \ref{dists}), we adopt a distance modulus of 27.70, equivalent to the average of the distance moduli from the four most recent published studies. We adopt a value of E(F606W-F814W)=0.16, obtained by converting the average E(B-V) value to E(F606W-F814W) using the Sirianni et al. (2005) relations. The morphology of the CMD is typical for a mixed stellar population with a range of ages and abundances. There is a dominant first ascent red giant branch (RGB) exhibiting an obvious termination at the core helium flash also known as the RGB tip at F814W $\approx$ 24. There appears to be a significant population of asymptotic giant branch (AGB) stars above the RGB tip signifying the presence of an intermediate-aged population. The CMD also shows a young main sequence at approximately zero color and a sequence of more massive brighter giants extending from F814W $\approx$ 21 down to F814W $\approx$ 25 where it merges with the RGB.

\subsection{Color Comparison}\label{Colors}

\begin{figure*}
  \centering
    \includegraphics[width=\textwidth]{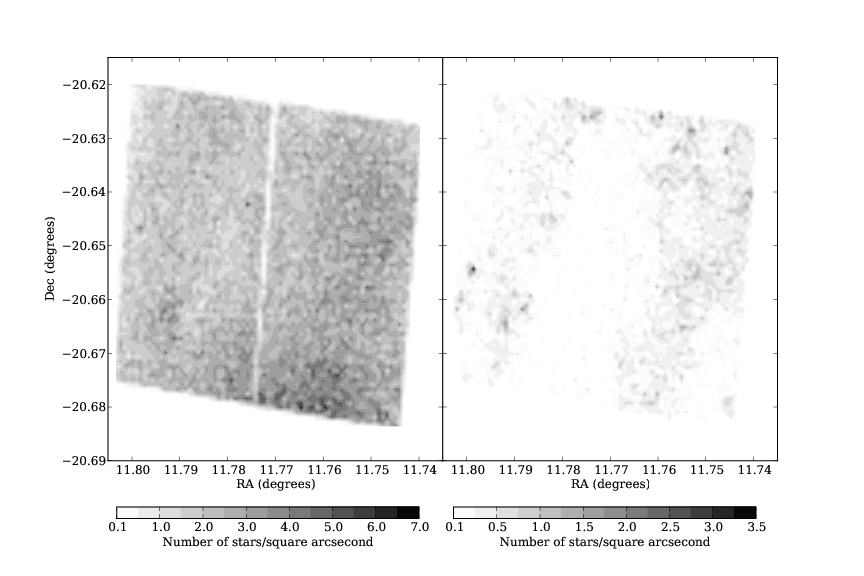}
  \caption{A comparison of the number density of red and blue stars in the void region, with darker regions being denser. The left panel shows stars with (F606W -- F814W) $\textless$ 0.2 while the right panel includes stars with (F606W -- F814W) $\textgreater$ 0.2. Only stars brighter than 26th magnitude in F814W and with (F606W-F814W) colors between --1 and 3 are shown. The red stars in our photometry show little variation in density across the void region. The blue stars, however, display a clear gradient in density across the void region.}
  \label{ColorComp}
\end{figure*}

We begin our analysis by comparing the locations and distributions of the bluer stars (comparatively younger) to those of the redder (and comparatively older) stars. Making cuts to our data, we assign ``red" stars as those with a F606W-F814W color greater than 0.2 and the ``blue" stars with colors less than 0.2, for all stars with F814W $\textless$ 26 and --1 $\textless$ (F606W -- F814W) $\textless$ 3. A comparison of the density distribution of these stars across the void region is seen in Figure \ref{ColorComp}, where the left panel shows the red stars while the right panel shows the blue stars.

We see that the red stars show little variation in density across the void region. In contrast, very few blue stars populate the area of the void and their density changes abruptly at the edge of the void. The relatively uniform density of the red stars, in conjunction with the observations from the UV to the IR also clearly showing the void region, imply that the appearance of the void is not simply due to extinction. The difference between the red and blue stellar densities suggests that the stellar population of NGC 247's void region is dominated by old stars, while the region just outside the void contains both old and young stars. We investigate this point further and in a quantitative manner in Section \ref{Ages}.

\subsection{Ages}\label{Ages}

\begin{figure*}
  \centering
    \includegraphics[width=\textwidth]{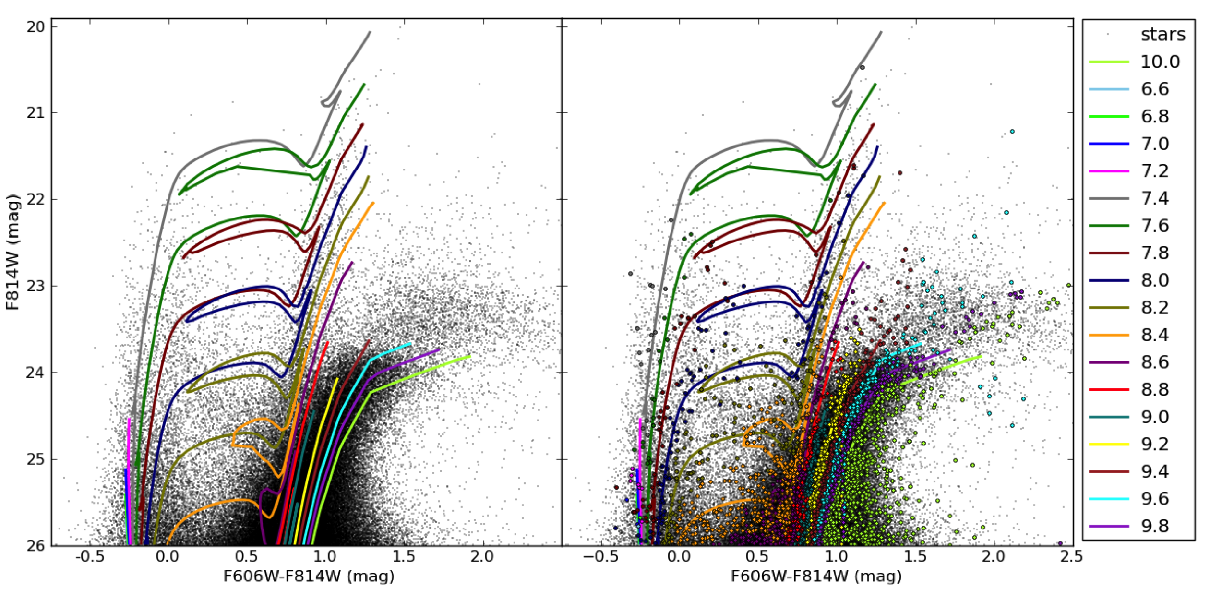}
  \caption{Left: The CMD of our photometry and the grid of isochrones. Right: The same as the left panel except that a randomly chosen subset of 3000 stars is overplotted on the CMD and isochrones. The colors of the plotted stars correspond to their attributed ages.}
  \label{ISOcheck}
\end{figure*}

A grid of Padova-Girardi isochrones was generated from log(t)=6.8 to log(t)=10.0, in steps of $\Delta$log(t)=0.2, and Z=0.004. We use this metallicity because at the adopted distance and reddening of NGC 247, the Z=0.004 tracks most closely match the location of the RGB. These isochrones are plotted along with the CMD in the left panel of Figure \ref{ISOcheck}. Using these isochrones as a comparison, a nearest neighbor interpolation was used to assign ages to stars in the CMD. 

The interpolation scheme determines where the star's color intersects each isochrone (this may happen up to three times, depending on the color and isochrone shape). At each of these intersections, a local linear interpolation is used to determine the magnitude of each isochrone at the star's color. The magnitude of each isochrone is compared to the actual observed magnitude of the star. The same process is repeated using the star's magnitude as a basis to predict the color which is then compared to the observed color. For each star, the isochrone that produces the lowest combined residual value of interpolated magnitude and interpolated color as compared to the star's actual observed magnitude and color is used to assign an age to each star.

We use this interpolation method to determine the ages for all stars in the CMD with F814W $\textless$ 26 and --1 $\textless$ (F606W -- F814W) $\textless$ 3. The result of this procedure is shown in the right panel of Figure \ref{ISOcheck}, where we plot a subsample of 3000 random stars with their assigned ages (noted by the associated color) along with the isochrones. These ages are used to analyze the properties inside and outside of the void region.

\begin{figure}
  \centering
    \includegraphics[scale=0.46]{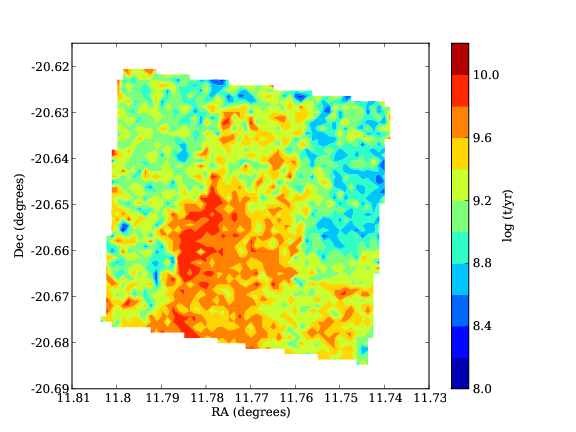}
  \caption{A contour map of ages across the void region in NGC 247. The median ages have a range of $\Delta$log(t)=0.2, from log(t)=8.0 to log(t)=10.0 as indicated by the color bar, with red being older and blue being younger.}
  \label{Contour}
\end{figure}

From the ages determined using the interpolation method described above, we map the median ages spatially across the void as seen in Figure \ref{Contour}. This figure shows a clear discontinuity in the relative ages of the stars across the void region in NGC 247. There are almost exclusively older stars in the void region with respect to those outside, indicating a severe deficiency of younger, recently formed stars in the void. This would suggest that star formation has halted in this area of the galaxy, but continues in the neighboring spiral arm.

\begin{figure}
  \centering
    \includegraphics[scale=0.5]{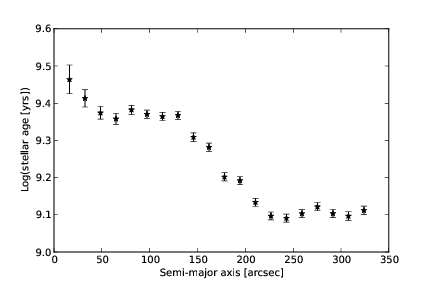}
  \caption{ The filled symbols represent the mean age of stars brighter than F814W = 26 in elliptical bins with increments of 16 arcseconds along the semi-major axis of the void. We see a clear drop in the mean age at 200 arcseconds (3.4 kpc) from the void center, suggesting that the stars outside of the void are a distinctly younger population than those inside.}
  \label{ageradius}
\end{figure}

In order to examine the relationship between the ages of the stars and their location, we have constructed a radial plot of mean age from the center of the void region. We used elliptical bins in increments of 16 arcseconds along the semi-major axis of the projected ellipse, starting from the approximate center of the void and moving outwards. The center of the void is approximately at an RA of 11.7730 and declination of --20.6515, with a position angle of 75.4 degrees. The result is 20 bins with at least 228 stars in each bin, and an average of 3351 stars per bin. In each bin, we determine the mean age of the stars and plot mean age as a function of semi-major axis in order to compare the ages of the stars inside the void to those outside. 

The result of this exercise is shown in Figure \ref{ageradius}, where we notice a dramatic result. The void region has a clear boundary around 200 arcseconds from the center ($\approx$3.4 kpc), beyond which there is a clear difference in the mean age of the population. This implies the void region is devoid of recent star formation unlike the region immediately surrounding it, which shows evidence for the presence of young stars.

\subsection{Star Formation Tracers}\label{Star Formation Tracers}

In addition to the photometry, which indicates that there has been a lack of star formation in the void region for some time, it is informative to examine other indicators of star formation in the disk and void of NGC 247. 

The HI distribution in NGC 247 suggests that there is gas in the northern region of the galaxy (Ott et al. 2012; Warren et al. 2012), though the brightness of the HI is very patchy across the entire disk. A visual comparison between the HI image from Ott et al. (2012) and Warren et al. (2012) indicates that the void area corresponds to a lower density region in the HI disk. However, we stress that the weaker HI emission in the void region is not unique, as areas of depressed HI flux are observed at many locations in the galactic disk. This patchy distribution of HI is typical in dwarf galaxies, likely due to dwarfs having shallower gravitational potential wells and thicker disks (Bagetakos et al. 2011, Roychowshury et al. 2010).
%The HI distribution in NGC 247 has many low-density regions and the low HI concentration in the void is not unique within the galaxy, which has HI holes throughout the disk. 

\begin{figure*}
  \centering
    \includegraphics[width=\textwidth]{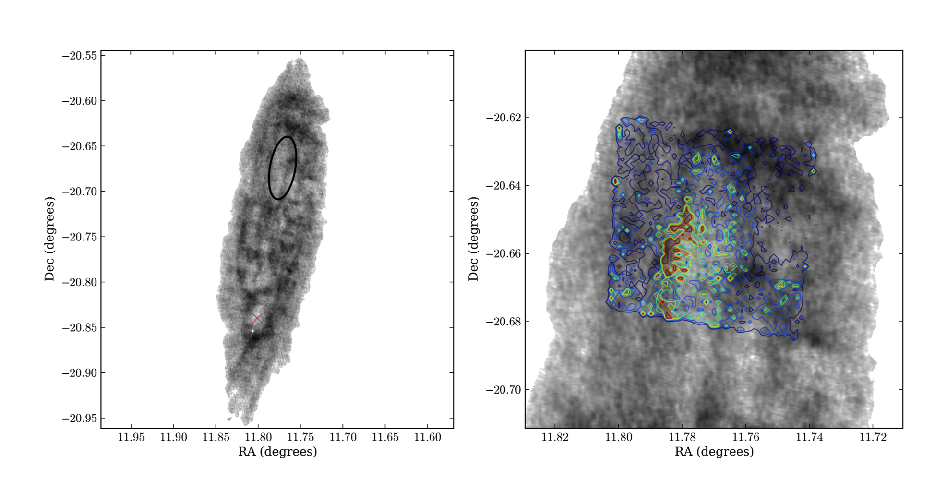}
  \caption{The HI emission in NGC 247 from Ott et al. (2012) and Warren et al. (2012). The left panel shows the HI emission in grey scale (darker is stronger emission) over the entire HI disk of NGC 247 with the location of the void indicated by the black ellipse. The red cross indicates a similar HI depression at an RA of 11.8000 and declination of --20.8389, used for comparison to the stellar void (see Figure \ref{HIplot}). The right panel again shows the HI emission, centered more closely on the void region. The contour of ages (Figure \ref{Contour}) is over plot in the right panel for comparison.}
  \label{HIcontour}
\end{figure*}

\begin{figure}
  \centering
    \includegraphics[scale=0.5]{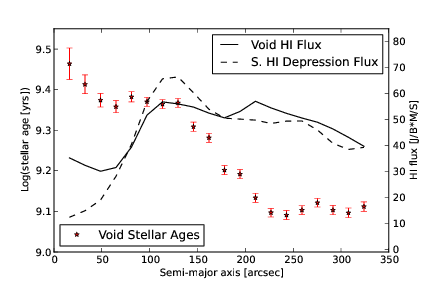}
  \caption{The derived stellar age gradient in the void overlaid with the HI flux profiles of two regions in the HI disk. HI flux profiles are constructed using annular bins of the same ellipticity as those used to find the mean stellar age profile of the void. The HI flux profile of the region coincident with the void and that of the southern HI depression (marked as a red cross on Figure \ref{HIcontour}, location at an RA of 11.8000 and declination of --20.8389) exhibit similar behavior. This suggests that the HI hole located in the void does not drive the observed stellar age gradient.}
  \label{HIplot}
\end{figure}

For a more quantitative comparison, we obtained the HI flux data from Ott et al. (2012) and Warren et al. (2012) to compare to our mean stellar ages in the void region. Figure \ref{HIcontour} shows the location of the void region in the galactic disk in the left panel, indicated by the black ellipse, and the HI emission with a contour of ages (Figure \ref{Contour}) overplotted on the void in the right panel. While the void does correspond with a region of depressed HI flux, it is not necessarily correlated with the behavior of the stellar age gradient in the void. To illustrate this, we found the HI flux profile of two visually similar regions, one at the location of the void and the other on the southern half of the disk, located at the red cross in the left panel of Figure \ref{HIcontour}. We determine the HI flux profiles by using annular bins spaced 16 arcseconds apart of the same ellipticity and location as those used to determine the radial stellar age gradient of the void. As seen in Figure \ref{HIplot}, the depression of HI flux coincident with the void region is not unique to the void, as both the HI flux profiles of the void region and the southern HI depression have similar profiles (the solid and dashed lines, respectively). Thus, although there is a correlation between HI emission and stellar age, the lack of HI is likely not a driver of the stellar age gradient present in the void, as the HI depression in the void region is not unique in the galaxy.

\begin{figure*}
  \centering
    \includegraphics[width=\textwidth]{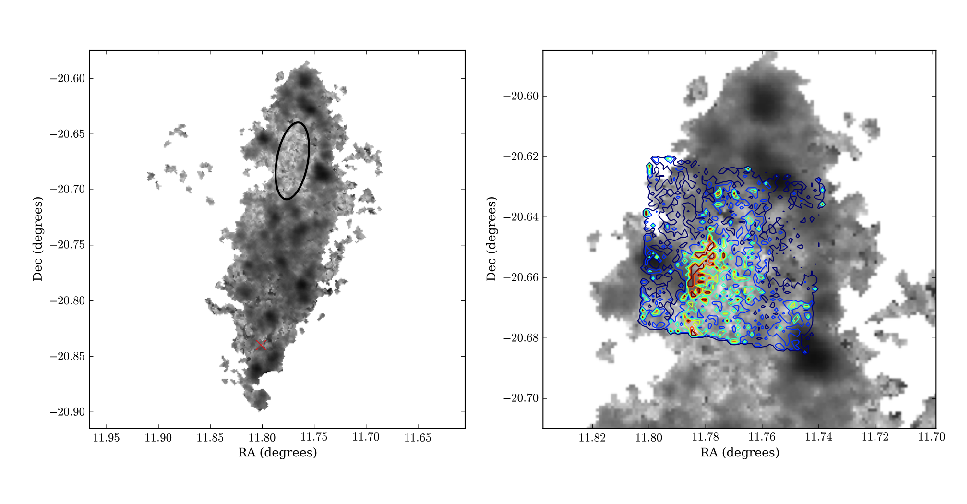}
  \caption{The left panel shows the H-$\alpha$ emission from Hlavacek-Larrondo et al. (2011) in grey scale across the disk. A black ellipse shows the location of the void of NGC 247 and the red cross indicates the location in the disk used for comparison to the stellar void (see Figure \ref{Halphaflux}, located at a right ascension of 11.7707 and a declination of --20.6714). The right panel shows the H-$\alpha$ emission again, centered more closely on the void region. The contour of ages (Figure \ref{Contour}) is also plot for comparison.}
  \label{Halpha}
\end{figure*}

H-$\alpha$ emission and HII regions are also key star formation tracers. While NGC 247 has sparse and relatively faint HII regions, the H-$\alpha$ emission has recently been mapped by Hlavacek-Larrondo et al. (2011). The middle left panel in Figure 3 of their paper shows the H-$\alpha$ emission in the disk of NGC 247. We plot the H-$\alpha$ emission across the disk in Figure \ref{Halpha}. The left panel shows the location of the H-alpha depression in the void region of the disk, indicated by the ellipse, and the right panel has the age contour (Figure \ref{Contour}) overplot on the H-$\alpha$ data from Hlavacek-Larrondo et al. (2011). 

\begin{figure}
  \centering
    \includegraphics[scale=0.5]{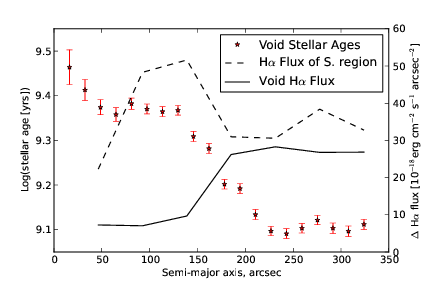}
  \caption{The derived stellar age gradient is overlaid with H-$\alpha$ flux profiles of two regions in the disk: one in the void and one in the opposite (southern) side of the disk. The H-$\alpha$ flux profiles are found using annular bins of the same ellipticity as those used to find the mean stellar age profile of the void. The H-$\alpha$ flux profile of the void region is shown as a solid line, and has a inverse correlation with the age moving outwards from the center of the void. We place the center of the elliptical bins to match the H-$\alpha$ depression in the void at a right ascension of 11.7707 and a declination of --20.6714. The dashed line shows the H-$\alpha$ flux profile of a southern region in the disk (indicated by the red cross in Figure \ref{Halpha} at an RA of 11.8000 and declination of --20.8389). The southern region does not show any such correlation with age and has a higher overall H-$\alpha$ flux than the void region. The extent and severity of depressed H-$\alpha$ emission in the northern region of NGC 247 is unique to the H-$\alpha$ disk and coincides with the location of the stellar void, suggesting a lack of recent star formation in that area of the disk.}
  \label{Halphaflux}
\end{figure}

Visual inspection shows a lack of H-$\alpha$ emission in the northern region corresponding to the location of the void, while the remainder of the disk has relatively mild to strong H-$\alpha$ emission. Unlike the HI disk, the decreased H-$\alpha$ emission of the Northern half of NGC 247's disk is unique and substantial, with a much lower overall H-$\alpha$ flux inside the void region compared to other parts of the disk. We have calculated the H-$\alpha$ flux profile in the region of the void in the manner described above with 12.9" elliptical bins and find that the H-$\alpha$ flux is minimal at the center of the void and grows steadily out to $\approx$200 arcseconds ($\approx$3.4 kpc) from the center of the void (see Figure \ref{Halphaflux}, solid line). As the ages of the stars get younger outwards from the center of the void, the H-$\alpha$ flux grows stronger.

We compare the distribution of H-$\alpha$ flux in the void region to the flux in a region on the opposite side of the disk (Figure \ref{Halphaflux}, dashed line). In the southern region of the disk, we find an H-$\alpha$ flux profile that shows no obvious trend. This region has a much higher H-$\alpha$ flux ($\sim$35$\times$10$^{-18}$ erg cm$^{-2}$ s$^{-1}$ arcsec$^{-2}$) than that of the void region within 200" ($\textless$10$\times$10$^{-18}$ erg cm$^{-2}$ s$^{-1}$ arcsec$^{-2}$). The lack of any significant H-$\alpha$ emission in the northern void region of NGC 247 compared to the rest of the disk again suggests a lack of recent star formation in that area.

CO observations could help to determine whether the HI depletion in the region of the void is significant in comparison to the rest of the HI disk, but unfortunately, these observations do not yet exist in sufficient resolution.

%%%%%%%%%%%%%%%%%%%%%%%%%%%%%%%%%%%%%%%%%%%  Discussion

\subsection{Discussion}\label{Discussion}

Other galaxies (such as NGC 1620; see Chaboyer \& Vader 1991, Vader \& Chaboyer 1995) have been found to host kiloparsec-sized holes. However, the void region in NGC 247 is several kiloparsecs in size and is a significant portion of the dwarf galaxy's disk. The analysis presented herein finds that the void region of NGC 247 exhibits a distinctly older population than the surrounding stars. Based on Figure \ref{Contour} and Figure \ref{ageradius}, we estimate that there has been very little star formation in the void for approximately the past Gyr. 

While ``bubbles" found in other galaxies have been attributed to multiple concurrent supernovae or gamma ray bursts, there are no supernovae remnants, X-ray sources, or large stellar clusters that have been detected within the void region in NGC 247 (Chaboyer \& Vader 1991; Vader \& Chaboyer 1995; Efremov et al. 1998; Jin et al. 2011; Tao et al. 2012). While this does not rule out the possibility that SNe or GRBs contributed to the formation of the void, there is no evidence to suggest that this is indeed the case in NGC 247.

Other star formation tracers, such as HI, HII, H-$\alpha$, and CO can be helpful in shedding light on the star formation history of NGC 247's void. We find a depression of HI flux in the void, but do not find that this is unique to that region, as there are similar depressions in the patchy HI disk. Thus, the HI is not likely to be a driving factor of the void region. The distribution of H-$\alpha$ emission is suggestive of a lack of recent star formation in the void. HII observations are dim and sparse; CO observations, if obtained, would be very useful in studying the star formation (or lack thereof) in this region.

One possibility for the seemingly long-lived nature of the void could be spiral density waves, which can provide an explanation for the existence and longevity of the void as part of a ``lop-sided" spiral. Although spiral density waves are often not very strong in dwarf galaxies (Ott et al. 2012, Brosch et al. 1998), interacting spiral density modes could lead to a discontinuity in the disk similar to that of NGC 247, as suggested by the simulations of Minchev et al. (2012). Alternatively, a transition from one spiral density pattern speed to another could also cause a discontinuity in a galaxy's morphology (Minchev et al. 2012). Both of these cases could produce a region empty of young stars in the galactic disk, similar to what we see in the void. However, we would expect the distribution of gas to follow the pattern of the discontinuity, and this is unequivocally not the case with the HI distribution. Additionally, the absence of a strong bar in NGC 247 makes a comparison to simulations by Minchev et al. (2012) incomplete, as their models are contingent on the presence of a strong bar. Nonetheless, spiral density waves provide a plausible scenario for the formation of the void, due to modal resonances and an explanation for a lifetime on the order of 1 Gyr.

Alternatively, we consider the possibility that an interaction with a dark-matter dominated sub-halo could cause the long-lived void. Recently, Kannan et al. (2012) showed that interactions with dark-matter dominated sub-halos that contain \emph{some} gas will lead to the formation of holes and shells in galactic disks.  Interactions with purely dark sub-halos lead to a localized high density region as the sub-halo gravitationally perturbs the disk (Chang \& Chakrabarti 2011; Chakrabarti et al. 2011).  However, such interactions with a purely \emph{collisionless} component cannot create holes in the gas distribution. This is due to the lack of contact forces, which is needed to push away the gas in the galactic disk.  The presence of a small amount of gas ($\sim$ few percent), i.e., a collisional component, will lead to the formation of holes, as the gas in the sub-halo can displace the gas distribution in the galactic disk.  Kannan et al. (2012) show that interactions with $\sim$ $10^{8}~M_{\odot}$ sub-halos with a gas fraction of $\sim$ 3 \% can create kpc-sized holes. Although we find the void in NGC 247 to be bigger than the largest holes of 2 kpc in their simulations, a higher gas fraction or velocity could plausibly account for a larger void. Moreover, the simulations of Kannan et al. (2012) indicate a deficit of new stars in the holes formed by gaseous sub-haloes and a corresponding increase in the density of new stars around the rim of the hole.  The reason for this is that a gaseous sub-halo that plunges through the disk produces a low density region surrounded by a high density wave.  This finding is consistent with our observational analysis of the ages of stars in and around the void region of NGC 247.  A deficit of new stars in HI holes and a concomitant enhancement in the star formation rate around the rims of holes is seen in other systems also.  For example, Pasquali et al. (2008) analyze the ages of stars in IC 2574 associated with HI holes to find a similar result as we do here.

The typical velocity perturbation due to an interaction with a sub-halo is $\sim$ $V_{\rm max}$, i.e., the maximum circular velocity, exerted over a length-scale that is comparable to the scale length of the sub-halo (Binney \& Tremaine 2008).  For impact parameters comparable to the scale length of the sub-halo, the size of the imprint is of the order of the scale length and drops for larger impact parameters.  The velocity dispersion of HI disks is $\sim 7~\rm km/s$ (Walter et al. 2008), and interactions with sub-halos that deliver more kinetic energy than the restoring energy provided by this velocity dispersion will leave an imprint in the disk.  We can use the scaling relations for halos in cosmological simulations (Maccio et al. 2008) to make some simple estimates of the sizes of holes.  If we take $M_{200} = 10^{8}~M_{\odot}$ (where $M_{200}$ is the mass out to the virial radius $R_{200}$, the radius where the density is 200 times the critical density), the sub-halo has a scale length of $\sim$ kpc for a concentration, $c \sim 18$, ($c = R_{200}/R_{\rm s} = 18$), where $R_{\rm s}$ is the scale length of the sub-halo.  Thus, close interactions with a $10^{8}~M_{\odot}$ sub-halo with a scale length $\sim$ kpc will lead to a localized high density region in the galactic disk if the sub-halo is purely dark, and to holes (similar to the $\sim$ kpc-sized hole in NGC 247) if the sub-halo contains some gas.    

The ALFALFA survey has recently uncovered a population of ultra-compact high velocity clouds (UCHVCs) (Giovanelli et al. 2010).  The discovery of optical emission from one of these UCHVCs enabled a distance and mass determination for the first time (Rhode et al. 2013).  The dynamical mass implied by these measurements suggests that some UCHVCs host a dark matter halo and may be ultra-faint dwarf galaxies.  Simulations by Keres \& Hernquist (2009) indicate that a large fraction of cold gaseous clouds can originate from ``cold-mode" accretion.  If so, structures similar to UCHVCs may be commonplace, and if a sufficiently massive UCHVC were to be on a perturbing orbit -- it would leave a detectable imprint (void) in the galactic disk. A high gas fraction sub-halo is also consistent with a sub-halo belonging to a population of UCHVC, as may be the case in NGC 247. Whether or not interactions with gas-bearing sub-halos can statistically reproduce the population of HI holes in galactic disks is a difficult question.  In this particular case however, there is a good correspondence between theoretical expectations and the observations.  Kannan et al.'s (2012) simulations show that gas-bearing sub-halos that are massive enough can displace the gas distribution in the galactic disk, leading to a deficit of new stars in the void and an enhancement of new stars close to the rim of the hole -- very similar to what we see for the void in NGC 247.  The lifetimes of holes depend on the orbital parameters of the sub-halo, ranging from 10 Myr to 70 Myr. Sub-halo interactions are not sufficient to account for all HI holes; smaller holes are likely to be due to occurrences of SNe and GRB activity while larger HI holes may be due to sub-halo interactions.

While many dwarf galaxies lack substantial internal shear (Ott et al. 2012, Weidner et al. 2011, Brosch et al. 1998), NGC 247's H-$\alpha$ rotation curve indicates significant shear across the void region (Hlavacek-Larrondo et al. 2011). The impinging gas-bearing sub-halo will displace some of the gas in the HI disk, and over a dynamical time, the void in the gaseous disk will be filled.  However, in order for that gas to form new stars, it has to be compressed to a high enough density to be Jeans unstable.  The large observed shear may prohibit gas compression.  A number of authors have noted an observed correlation between large shear rates and low star formation (Seigar 2005; Leroy et al. 2008).  Deep, high resolution CO observations would provide a critical test of this scenario.

Features in a collisionless system may be phase mixed over many dynamical times, but a time-scale for this process cannot be defined in generality.  The rate of phase mixing is inversely proportional to the maximum and minimum  orbital frequencies considered (Merritt \& Valluri 1996), and can be of order many dynamical times.  In some potentials (such as triaxial potentials), the rate of mixing can be enhanced by chaotic mixing, but stochastic trajectories are very sensitive to initial conditions (Merritt 2013).  Triaxial potentials are expected in systems that have undergone mergers (Allgood et al. 2006), and the non-Keplerian motion indicated by the observed rotation curve of NGC 247 (Hlavacek-Larrondo et al. 2011) may be reflective of a complex potential.  While a timescale for features in the stellar disk, such as the void, is difficult to establish in generality, one clear prediction of our model is a corresponding lack of CO in the void region.

%%%%%%%%%%%%%%%%%%%%%%%%%%%%%%%%%%%%%%%%%%%  Conclusion

\section{Conclusion}\label{Conclusion}

In this work, we have reduced and analyzed archival HST imaging of the void region in the Sculptor group dwarf galaxy NGC 247. Based on our deep color-magnitude diagram for this region, we find that the stars inside the void of NGC 247 are significantly older than those on the edge and outside of the void. This suggests a lack of star formation in recent epochs inside the void. This conclusion is supported by H-$\alpha$ observations from Hlavacek-Larrondo et al. (2011) and examination of the stellar populations inside and outside the void region. 

From our analysis of NGC 247, we present the following specific conclusions:

1. We estimate the size of the void to be approximately 3.4 kpc long at a distance modulus of 27.7. This is a large portion of the 11.34 kpc major axis of the galaxy.

2. The stellar population inside the void is fundamentally different than that outside the void region. The mean stellar age outside the void is approximately 9.1, while the mean stellar age of stars close to the void center is 9.4. Our analysis of age vs. radius suggests little to no star formation in the void for more than 1 Gyr.

3. We find the characteristics of the void are inconsistent with typical disk evolution of a galaxy. Star formation is suppressed in the void region of the galactic disk where we would usually find a more diverse stellar population. There is no evidence to suggest scenarios of SNe or GRBs creating this disruption or a strong, unique anti-correlation with HI flux. H-$\alpha$ emission also suggests a lack of young, hot stars being formed in the void region.

4. A $\sim$ kpc-sized hole with the observed age distribution may have been formed due to a close interaction with a $\sim 10^{8}~M_{\odot}$ gas-bearing sub-halo, as suggested by the recent simulations of Kannan et al. (2012). We would expect gas displaced by such an interaction with a gas-bearing sub-halo to repopulate the void region in approximately a dynamical time. However, shear in the void (as evidenced by the rotation curve) could prevent gas compression necessary for star formation, leading to a long-lived void. CO observations can be a crucial test of this scenario.

\section*{Acknowledgments}

We thank the anonymous referee for his/her comments which contributed to the improvement of the paper. We would also like to extend our gratitude to Rahul Kannan and Riccardo Giovanelli for their comments on an earlier version of this paper. We would like to thank Daniel Gettings for his generosity in the use and patient explanation of his elegant code as part of this project. We would also like to thank Julie Hlavacek-Larrondo for her willingness to provide us with her H-$\alpha$ observations. Additionally, we are grateful to Wanggi Lim for his initial contribution to developing this project.

\label{lastpage}

\end{document}